**Structure and physical properties of $Na_xCoO_2{\cdot}yH_2O$ superconducting system**

Y.G.. Shi[1], J.Q. Li[2], H. C. Yu[1], Y. Q. Zhou[2], H. R. Zhang[1], and C. Dong[1]

[1] National Laboratory for Superconductivity, Institute of Physics, Chinese Academy of Sciences, Beijing, People's Republic of China

[2] Beijing Laboratory of Electron Microscopy, Institute of Physics, Chinese Academy of Sciences, Beijing, People's Republic of China

The structural features and physical properties of $Na_xCoO_2$ and $Na_xCoO_2{\cdot}yH_2O$ materials have been investigated. The $Na_xCoO_2{\cdot}yH_2O$ samples in general undergo superconducting transitions at around 3.5K. EDAX analyses suggest our samples have the average compositions of $Na_{0.65}CoO_2$ for the parent compounds and $Na_{0.26}CoO_2{\cdot}yH_2O$ for the superconducting oxyhydrates. TEM observation reveals a new superstructure with wave vector q=<1/2,0, 0> in the parent material. This superstructure becomes very weak in the superconducting samples. Electron-energy-loss spectra (EELS) analyses show that the Co ions have the valence states of around +3.3 in $Na_{0.65}CoO_2$ and around +3.7 in $Na_{0.26}CoO_2{\cdot}yH_2O$.

The layered sodium cobalt oxyhydrate materials ($Na_xCoO_2{\cdot}yH_2O$) have attracted considerable interest because of its similarity to the high-Tc superconductors. Systematically theoretical and experiential investigations on this new superconducting system are expected to shed light on the superconducting mechanism in the High-Tc cuprates[1-6]. Like the high-Tc superconductors, the $Na_xCoO_2{\cdot}yH_2O$ crystal structure consists of electronically active planes (edge sharing $CoO_6$ octahedra) separated by (Na-$H_2O$) layers that act as charge reservoirs [3]. The Na content can be varied in the charge reservoir layers, which results in the same type of out-of-plane chemical doping control of in-plane electronic charge as found for the cuprate superconductors. It has been confirmed that the structural and compositional alternations in this layered system could yield evidently change of superconductivity [3,7,8]. Hence, a careful structural analysis, especially the evolution of microstructure along with the oxyhydration, should play an important role for understanding the significant properties of this new material. In present study, we report on the developments of our investigations on the structural and physical properties of the parent $Na_xCoO_2$ materials and the $Na_xCoO_2{\cdot}yH_2O$ superconductors, the valence states of Co ions in several specific materials, as measurement by EELS, have been briefly discussed.

The polycrystalline samples of $Na_{0.65}CoO_2$ were prepared by conventional solid-state reactions [9]. Superconducting $Na_xCoO_2{\cdot}yH_2O$ materials were prepared by $Na_{0.65}CoO_2$ in the excessive bromine solved in acetonitrile at ambient temperature for two to five days to deintercalate sodium, the detailed process and treatments in sample preparation are similar with that reported previously in ref. 1. The product materials were washed several times with acetonictrile and water, and stored in a humidified atmosphere for two days. Then it was pressed into pellets and protected in liquid nitrogen. Specimens for transmission-electron microscopy (TEM) observations were polished mechanically with a Gatan polisher to a thickness of around 50μm and then ion-milled by a Gatan-691 PIPS ion miller. In addition, we also prepared some thin samples for electron diffraction experiments simply by crushing the bulk material into fine fragments, which were then supported by a copper grid coated with a thin carbon film. The TEM investigations were performed on a H-9000NA TEM operating at the voltage of 300kV and a Tecnai F20 (200 kV) electron microscope with an atomic resolution of about 0.23nm. The EELS measurements were performed on the Tecnai F20 transmission electron microscope (equipped with a post column Gatan imaging filter). The energy resolution in the EELS spectra is 0.7 eV under normal operation conditions. In order to minimize the radiation damage under electron beam the samples were cooled below 200K during our TEM observations.

The basic crystal structures of the parent materials $Na_xCoO_2$ and the superconducting materials $Na_xCoO_2 \cdot yH_2O$ have been measured by x-ray diffraction (XRD). Figure 1(a) shows a XRD pattern obtained from a parent $Na_xCoO_2$ sample, all diffraction peaks in this pattern can be well indexed by an hexagonal cell with lattice parameter a=2.839Å and c=10.804Å. Figure 1(b) shows a XRD data obtained from a typical superconducting sample of $Na_xCoO_2 \cdot yH_2O$ with Tc=3.5K. It can be recognized that the crystal lattice has a longer c axis in the superconducting samples, which results from the intercalation of additional $H_2O$ sheets between $CoO_2$ layers. The basic parameters for the superconducting crystal is a=2.821 Å and c=19.807Å with the space group of $P6_3/mmc$. After careful analysis, we also find some weak diffraction peaks in the diffraction pattern from some other impurity phases.

Figure 2 shows the zero-field cooling DC magnetization data measured in a field of 20Oe for selected samples prepared under slightly different conditions. The superconducting transition occurs at around 3.5K for both samples. Strong diamagnetic signals appearing in these measurements provide direct evidence for bulk superconductivity in $Na_xCoO_2 \cdot yH_2O$ samples. Alternations of Na concentration from one area to another and the resultant structural inhomogeneities are likely to be the essential causes for rounding of the superconducting transitions.

Microstructure features of sodium cobalt oxyhydrate materials depend essentially on the synthesis process. Figure 3(a)-(d) show the scanning electron microscopy (SEM) images illustrating the typical microstructure of the $Na_{0.65}CoO_2$ and $Na_xCoO_2 \cdot yH_2O$ samples, both materials show up clearly layered structural features. The grain sizes range from 0.1μm to10μm in the $Na_{0.65}CoO_2$ materials and from 3μm to 30μm in the superconducting samples. The crystal structure of this kind of materials is based on the close packed layers of the edge-sharing $CoO_2$ octahedra perpendicular to the c axis, separated by intercaslent layers consisting of Na and $H_2O$ sheets, the crystals could be cleaved easily between $CoO_2$ layers and give rise to noticeable layered crystalline piece as displayed in the SEM images of figure 3 (c) and (d). In the superconducting materials, we in general can find the perfect crystal with the thickness of less than 1μm along c-direction. We also noted that the superconducting crystals are very unstable under electron beam illumination. Lowering the sample temperature down to 200K by using low-temperature stage can effectively prevent the crystal modification from the decomposition of $H_2O$ molecules. EDAX analyses have been used to measure the compositions of the crystal grains prepared under different conditions. The composition of the parent materials is estimated as $Na_{0.65}CoO_2$, and in some case it is likely that additional oxygen, intercalated among the structural layers, could be detected. In the superconducting materials, the EDAX measurements revealed that the Na concentration could change slowly from one grain to another. The composition is estimated in the range $Na_{0.25}CoO_2 \cdot yH_2O$ to $Na_{0.3}CoO_2 \cdot yH_2O$ for our superconducting samples. Moreover, the EDAX analysis confirms the presence of a few impurity phases in the superconducting materials, which is consistent with the x-ray diffraction results.

TEM observations reveal that the crystal structures, for both the parent compound and the superconducting phase, have the hexagonal lattice with the space group of P63/mmc in assistance with the reported data. The basic properties of the crystal structure can be clearly illustrated by convergent-beam electron diffraction patterns and TEM images obtained along several relevant zone-axis directions. Figure 4(a) shows the [001] zone CBED pattern, illustrating the 6-mm symmetry with the systematical mirror-planes located in the {100} and {210} crystal planes. A sixfold axis is along the c direction. The most strike feature revealed in our TEM observation is the presence of a superstructure within the ab crystal plane. This superstructure is strong and clearly visible in the parent $Na_{0.65}CoO_2$ material, and becomes weaker, even invisible, in the $Na_{0.3}CoO_2 \cdot yH_2O$ superconducting materials. Figure 4(b) shows an electron diffraction pattern taken from the parent material $Na_{0.65}CoO_2$, exhibiting the superstructure spots at the systematic (h+1/2, k, l) positions. Detailed analyses suggest that this superstructure results possibly from the mismatch between the $CoO_2$ layer and the $Na(H_2O)$ sheets. On the other hand, this superstructure is likely to be in correlation with chemical inhomogeneity commonly occurring in this kind of materials, e.g. the alteration of the Na concentration. *In situ* cooling TEM observation indicates that this superstructure is very stable within the temperature range from room temperature down to 100K. no structural phase transition is observed. A further investigation on the origin of this superstructure, by means of x-ray diffraction and high-resolution electron microscopy, is still under progress.

Figure 4 (c) shows a high-resolution electron micrograph of a $Na_xCoO_2{\cdot}yH_2O$ crystal taken along the [001] zone-axis direction, the superstructure fringe with the space of $2d_{100}$ are indicated by arrows. It is commonly observed that the areas with clear superstructure show up complex domain structure corresponding to different orientation variants. In the superconducting samples this superstructure becomes very weak, as a result, only the hexagonal structure feature of the sublattice is revealed in the high resolution TEM images, as shown in figure 4(d).

EELS analyses have been performed on both the parent material $Na_{0.65}CoO_2$ and superconducting materials $Na_xCoO_2{\cdot}yH_2O$. Figure 5(a) shows an EELS spectrum of a superconducting crystal taken from an area of about 100nm in diameter. In this spectrum the typical peaks, i.e. the collective plasmon peak as well as core edges for Co, O and Na elements, are displayed.

EELS, a powerful technique for material characterization at a nanometer spatial resolution, has been widely used in chemical micro-analysis [10]. For transition metals with unoccupied 3d states, the transition of an electron from 2p state to 3d levels leads to the formation of white lines. The $L_3$ and $L_2$ lines are the transitions from $2p_{3/2}$ to $3d_{3/2}3d_{5/2}$ and from $2p_{1/2}$ to $3d_{3/2}$, respectively. Their intensities are related to the unoccupied states in the 3d bands. In $Na_{0.3}CoO_2{\cdot}yH_2O$ materials, we have made a series of measurements in association with quantitative analyses by the method as reported by Wang et al [10]. Figure 5(b) shows a typical spectrum for the Co $L_2$ and $L_3$ peaks obtained from an area of about 20nm in size, in which we schematically illustrate the extraction of the intensities for the white lines of Co for a superconducting sample. As pointed out in the previous literatures, the ratio of $L_3/L_2$ is very sensitive to the valence state of Co[10, 11], a series of relevant experimental data and some well-established empirical models are also reported in the studies of the related materials. These results provide a basis for our measuring the Co valence in $Na_xCoO_2{\cdot}yH_2O$ materials.

Our first analysis was performed on the parent sample $Na_{0.65}CoO_2$, the results indicate that the ratio of $L_3/L_2$ in general is around 2.4, which could yield the Co valence of about 3.3-3.4 in the parent sample. The analyses of the superconducting sample give rise to $L_3/L_2$ at around 2.2, this data could change slightly from one grain to another. Our systematical analyses conclude that the Co valence state ranges from 3.6 to 3.8 for superconducting samples, this result is in good agreement with the data from other experiments [4-6].

In summary, we have prepared a series of the layered sodium cobalt oxyhydrates with evident superconducting transitions at around 3.5K. SEM observations clearly show the layered structure feature of this kind of materials. TEM investigation reveals a superstructure with wave vector q=<1/2,0, 0> appearing in the parent materials, which becomes weak and even invisible in the superconducting samples. Analyses of EELS on Co valence states suggest that the Co ions have the valence states ranging from +3.3 to +3.4 in the $Na_{0.65}CoO_2$ materials and from +3.6 to +3.8 in the superconducting materials.

Acknowledgments

We would like to thank Mr. W. W. Huang and Miss. S. L. Jia for their assistance in preparing samples and measuring some physical properties. The work reported here was supported by the ‘‘Hundreds of Talents’’ program organized by the Chinese Academy of Sciences, P. R. China, and by the ‘‘Outstanding Youth Fund (J. Q. Li)” with Grant No. 10225415.

References

[1]. Takada, K., Sakurai, H., Takayama-Muromachi, E., Izumi, F., Dilanian, R.A., Sasaki T., Nature **422**, 53 (2003).

[2]. Bednorz, J. G. & Müller, K. A., Z. Phys. B 64, 189 (1986)

[3]. Foo, M.L., Schaak, R.E., Miller, V.L., Klimczuk, T., Rogado, N.S., Wang, Yayu, Lau, G.C., Craley, C., Zandbergen, H.W., Ong, N.P, and Cava, R.J., Sol. St. Comm., in press.

[4]. Sakurai, H., Takada, K., Yoshii, S., Sasaki, T., Kindo, K.,Takayama-Muromachi, E., cond-mat/0304503

[5]. Baskaran, G., condmat/0303649

[6]. Kumar, B., and Shastry, B.S. condmat/0304210

[7]. Honercamp, Carsten., cond-mat/0304460

[8]. Tanaka, A., and Hu, X., cond-mat/0304409

[9]. Fouassier,C., Matejka, Gn., Reau, J.-M., and Hagenmuller, P. J., Sol St. Chem. 6, 532 (1973)

[10]. Wang, Z. L., Yin, S., Jiang,Y. D., Micron. 31, 571 (2000)

[11]. Pearson, D. H., Ahn, C. C, and Fultz, B., Physical Review B. 47, 8471 (1993)

Figure captions

Figure 1. Powder x-ray diffraction patterns of (a) $Na_{0.65}CoO_2$.and (b) $Na_xCoO_2{\cdot}yH_2O$.

Figure 2. Magnetic susceptibility curves of (a) $Na_xCoO_2{\cdot}yH_2O$ and (b)$Na_xCoO_2{\cdot}yH_2O$ protected in a humidified atmosphere for three more days.

Figure 3. SEM images showing the microstructure features of (a), (b) $Na_{0.65}CoO_2$.and (c), (d) $Na_xCoO_2{\cdot}yH_2O$, Layered structural feature are clearly illustrated in figure 3(b) and (d).

Figure 4(a) CBED pattern of $Na_xCoO_2{\cdot}yH_2O$ along c-axis direction, showing the 6mm symmetry. (b) Electron diffraction pattern showing the presence of superstructure in $Na_{0.65}CoO_{2.}$ (c) High resolution TEM image from an area with clear superstructure fringes. (d) High-resolution TEM image showing the hexagonal sublattice for the superconducting phase.

Figure 5 (a) EELS spectrum of the $Na_xCoO_2{\cdot}yH_2O$ superconductor showing the zero-loss peak, plasmon resonance and ionization edges arising from the elements of O, Co and Na. (b) An EELS spectrum acquired from a crystal in the superconducting sample, schematically illustration the method used to extract the intensities of white lines.

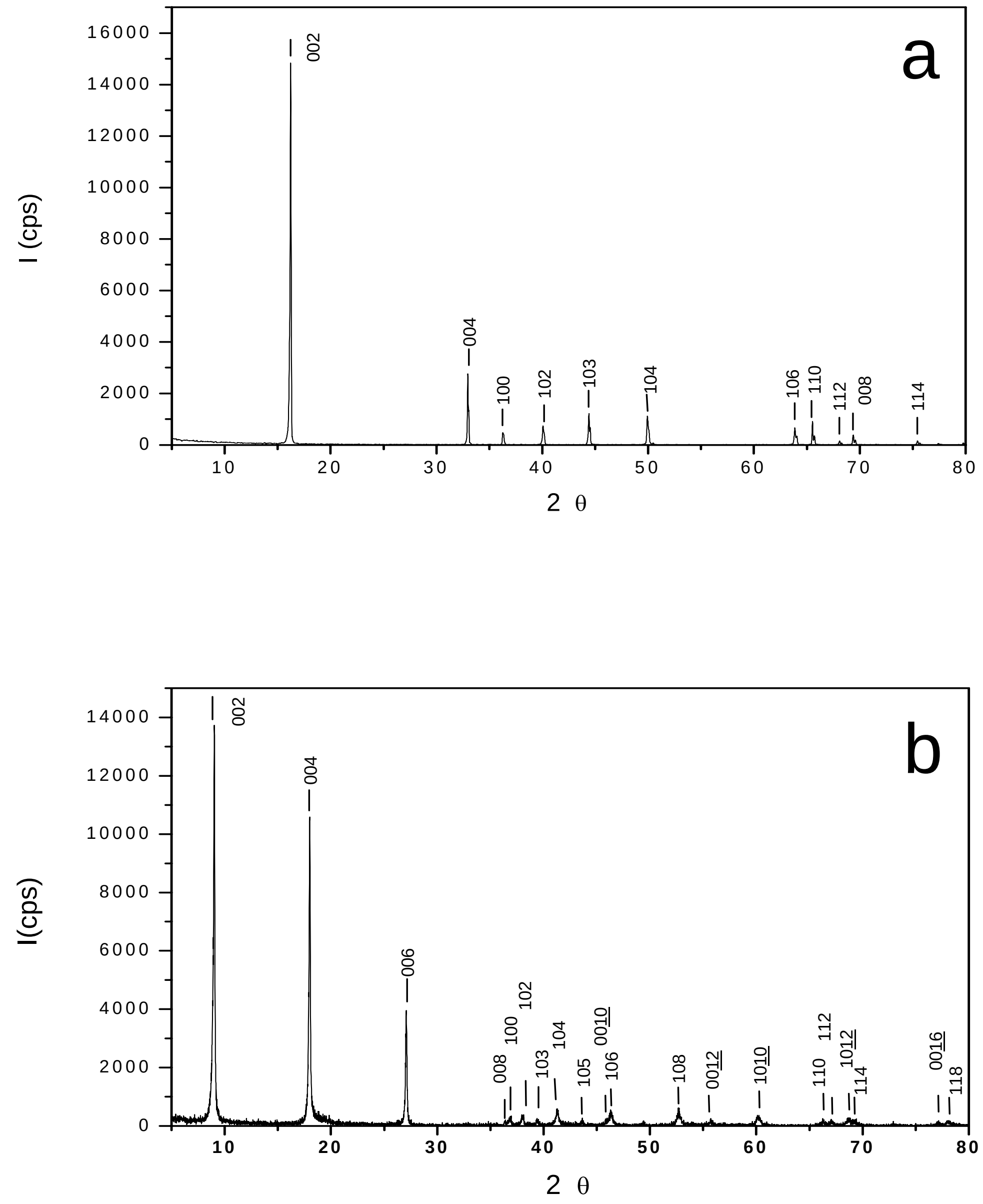

Figure 1

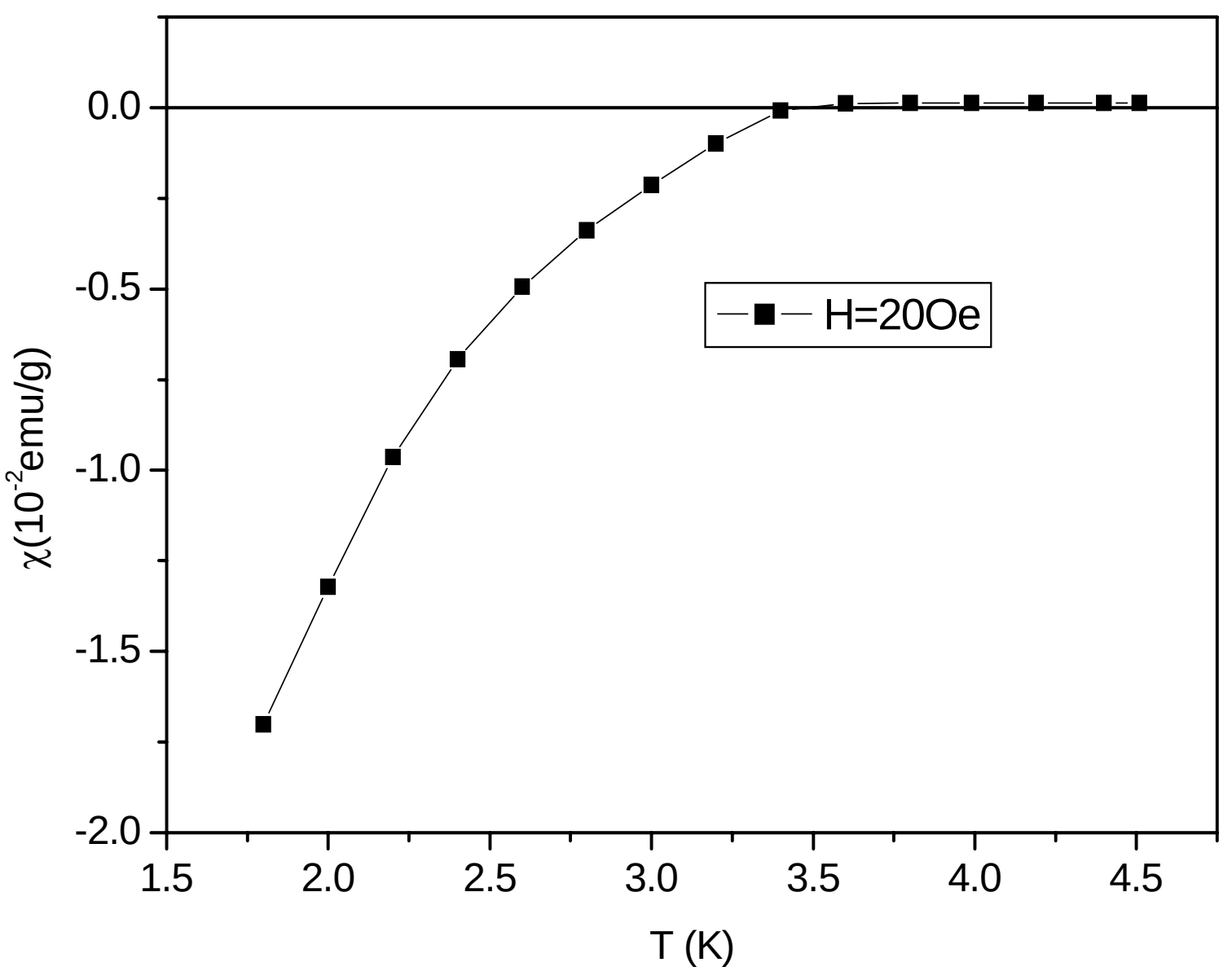

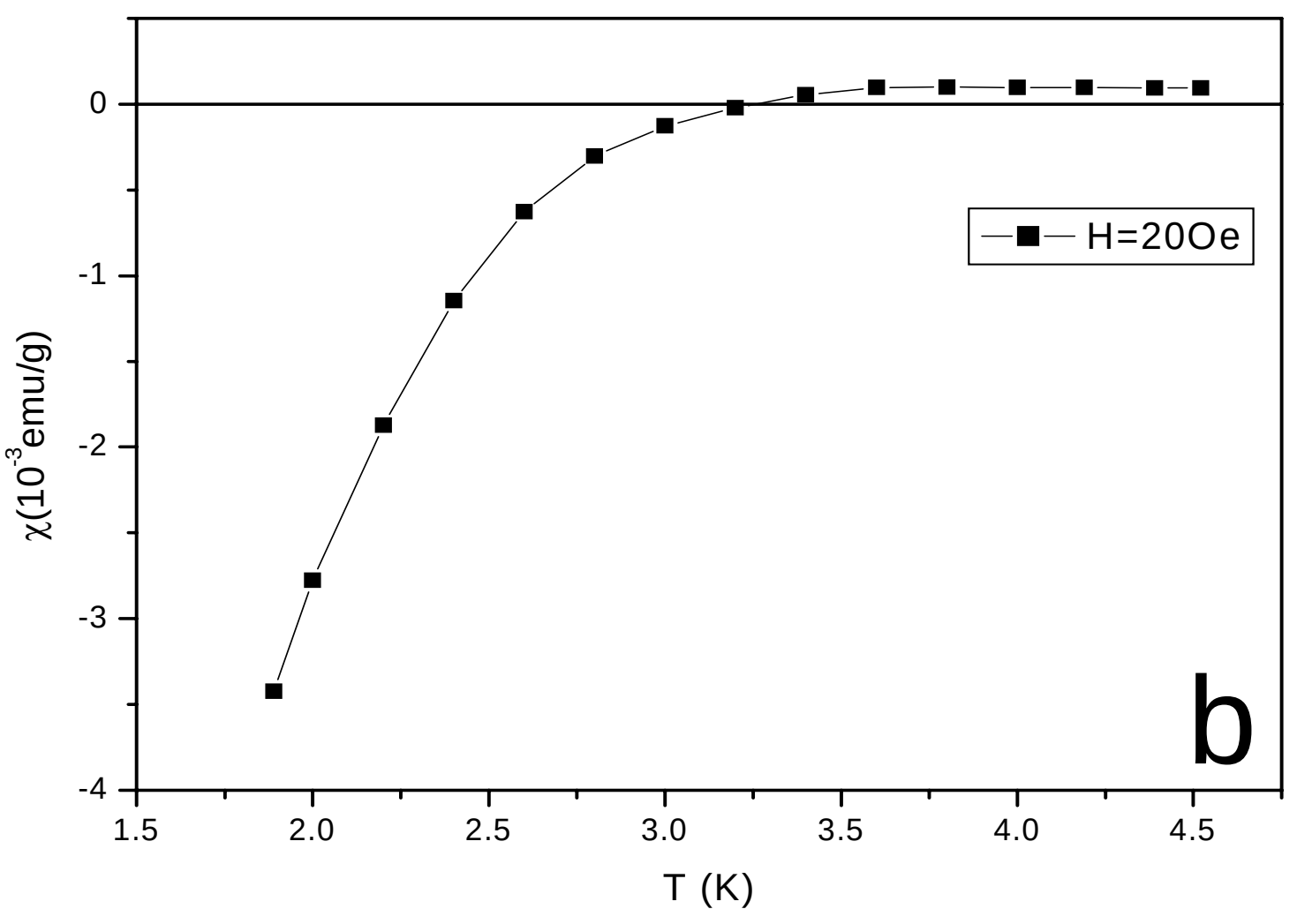

Figure 2

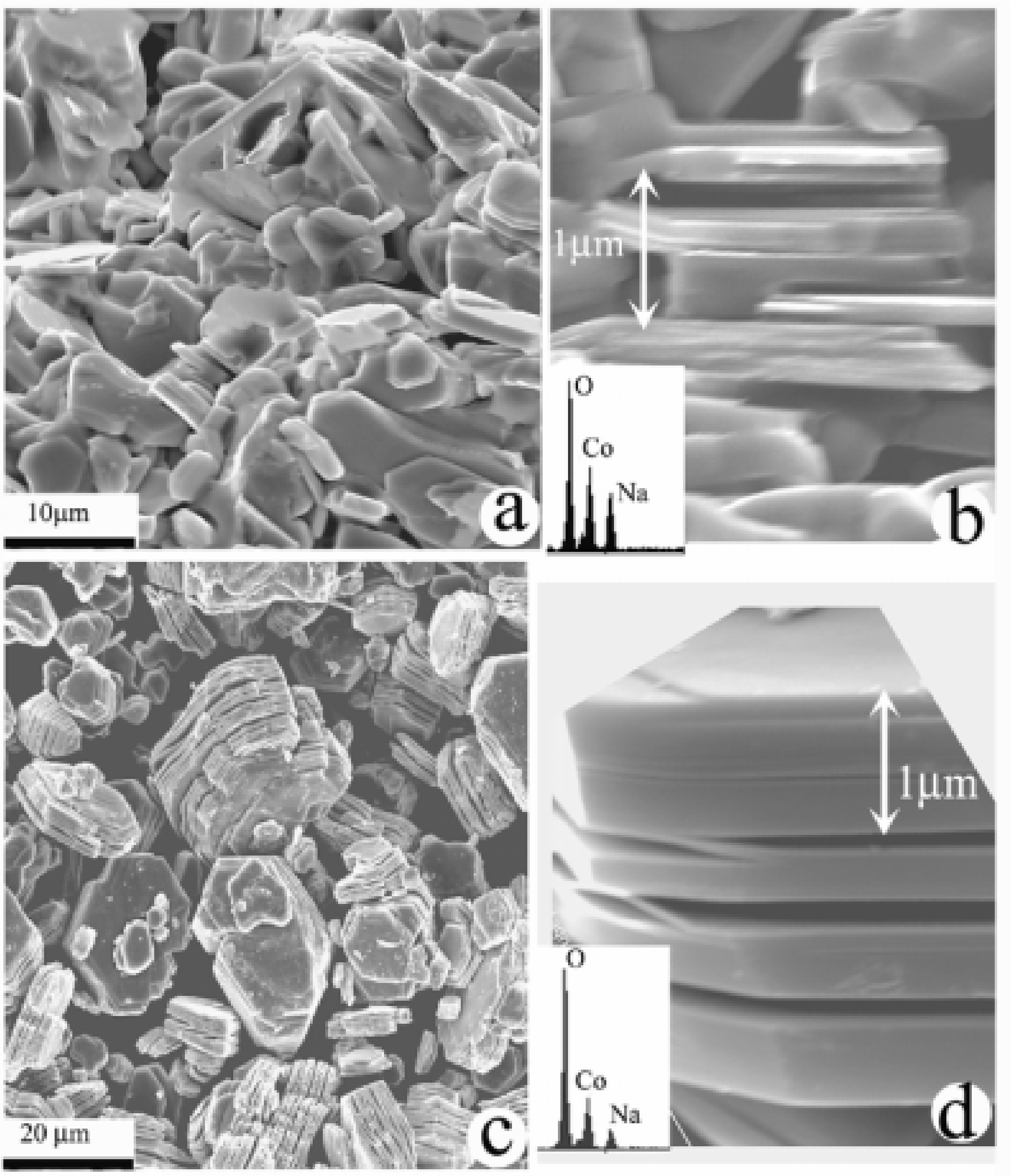

Figure 3

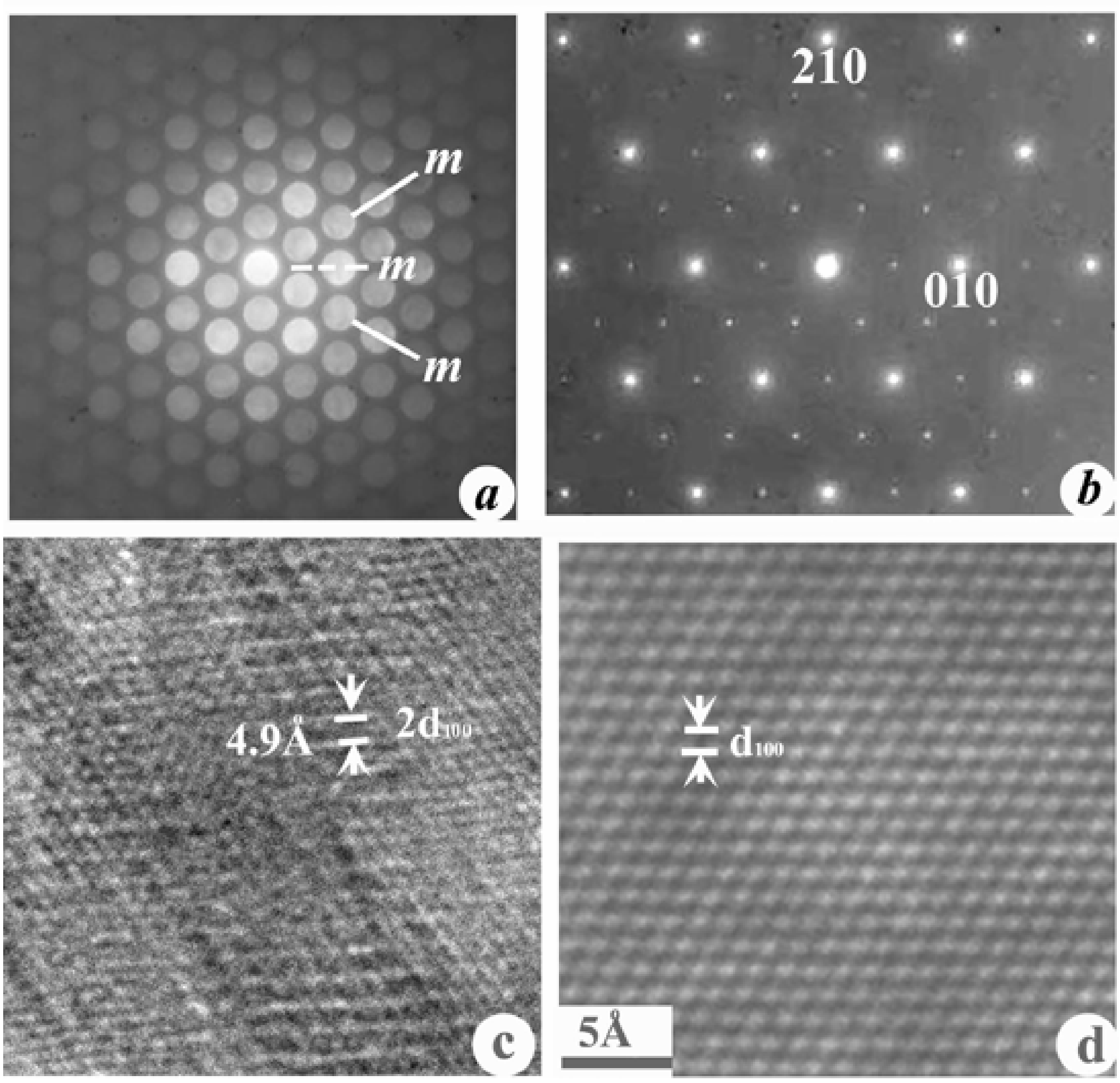

Figure 4

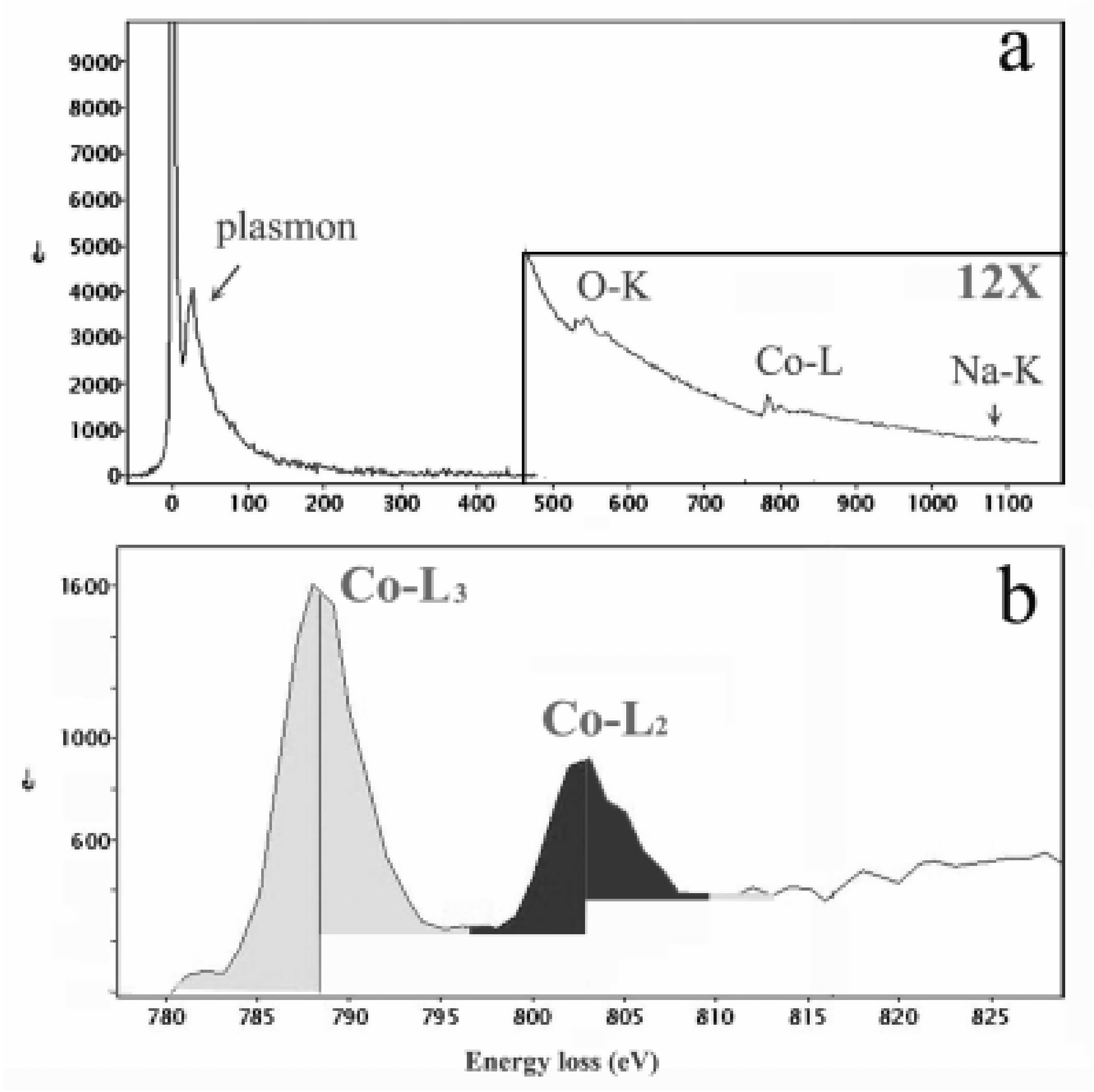

Figure 5